%% file: main.tex
\documentclass[a4paper,11pt]{article}
\usepackage{jinstpub} %
\usepackage{caption}
\DeclareUnicodeCharacter{2212}{-}
\usepackage{lipsum} 
\usepackage{siunitx}
\usepackage{amsmath}
\usepackage{xcolor}
\usepackage{subfig}
\usepackage{comment}
\usepackage{isotope}
\usepackage{ragged2e}

\usepackage{glossaries}
\glsdisablehyper
\input{acronyms}

\title{\boldmath Proto-0: a prototype for validating key technologies of the DarkSide-20k experiment and beyond}

\collaboration[c]{on behalf of the DarkSide-20k Collaboration}

\author[a]{R.~de~Asmundis,}
\author[b,a]{R.~Calabrese,}
\author[c,d]{M.~Caravati,}
\author[b,a]{G.~Fiorillo,}
\author[b,a]{L.~Flores,}
\author[a]{G.~Grauso,}
\author[a,1]{G.~Matteucci,\note{Corresponding author.}}
\author[e]{N.~Pino,}
\author[b,a,f]{D.~Rudik,}
\author[g,h,i]{M.~A.~Sabia,}
\author[b,a,j]{Y.~Suvorov}

\affiliation[a]{INFN Sezione di Napoli,\\
Via Cintia 21, Napoli 80126, Italy}

\affiliation[b]{Physics Department, Università degli Studi Federico II di Napoli,\\
Via Cintia 21, Napoli 80126, Italy}

\affiliation[c]{Gran Sasso Science Institute,\\
L’Aquila 67100, Italy}

\affiliation[d]{INFN Laboratori Nazionali del Gran Sasso,\\
Assergi (AQ) 67100, Italy}

\affiliation[e]{INFN Laboratori Nazionali del Sud,\\
Catania  95125, Italy}

\affiliation[f]{National Research Nuclear University MEPhI,\\
Moscow 115409, Russia}

\affiliation[g]{Physics Department, Sapienza Università di Roma,\\
Roma 00185, Italy}

\affiliation[h]{INFN Sezione di Roma,\\
Roma 00185, Italy}

\affiliation[i]{AstroCeNT, Nicolaus Copernicus Astronomical Center,\\
Polish Academy of Sciences, 00-614 Warsaw, Poland}

\affiliation[j]{National Research Centre Kurchatov Institute,\\
Moscow 123182, Russia}

\emailAdd{giuseppe.matteucci@infn.it}

\abstract{The DarkSide-20k experiment, currently under construction at LNGS, will employ a next-generation dual-phase liquid-argon Time Projection Chamber (TPC) with SiPM-based Photon Detector Units and low-background materials to achieve the ambitious goal of operating with an instrumental background close to zero.
Proto-0 is a small-scale dual-phase argon TPC operated at INFN Naples, designed to validate the integration of DarkSide-20k key technologies in a realistic detector environment and to study charge extraction and electroluminescence signal formation.
In this short paper we report on the early operation of Proto-0 and its single-phase commissioning. We focus on the measurement of the scintillation light yield using external and internal calibration sources.}

\keywords{Dark Matter detectors (WIMPs, axions, etc), Noble liquid detectors (scintillation, ionization, double-phase), Time projection Chambers (TPC), Detector design and construction technologies and materials}

\begin{document}
\maketitle
\flushbottom

\section{The DarkSide-20k Experiment}
\justifying

Weakly Interacting Massive Particles (WIMPs) are a promising candidate for a dark matter (DM) particle \cite{Roszkowski:2004jc}, and several experiments have attempted their direct detection over the last decades, yet without success. As the expected signal comes from rare events producing low-energy nuclear recoils, the challenge for such experiments is to construct an experimental apparatus capable of high exposure, low threshold and ultra-low background. One technology has stood out over the others for its background mitigation capabilities and scalability: the noble-element dual-phase Time Projection Chamber (TPC). Xenon and argon TPCs currently set the most stringent limits on WIMP interactions over a broad mass range \cite{Billard:2021uyg} and are planned to reach the limit of their sensitivity, the so-called neutrino fog \cite{OHare:2021utq}, at the next generation of detectors.

In a dual-phase TPC, particle interactions in the liquid target produce both prompt scintillation light (S1) and ionization electrons; the latter are drifted to the liquid–gas interface, extracted into the gas phase, and converted into a secondary scintillation signal (S2) proportional to the ionization charge. The simultaneous measurement of S1 and S2 provides event reconstruction and background discrimination capabilities. In this context, liquid argon detectors have proven to be extremely effective in background rejection, owing to the powerful pulse-shape discrimination techniques enabled by the distinctive scintillation properties of argon \cite{Matteucci:2024foy}. This was demonstrated by the DarkSide-50 experiment and its zero background result \cite{DarkSide:2018kuk}. Building on these capabilities, the DarkSide-20k Collaboration has designed and is currently constructing at LNGS a massive argon detector capable of reaching the neutrino fog.

\begin{figure}[t]
    \centering
    \includegraphics[height=0.40\linewidth]{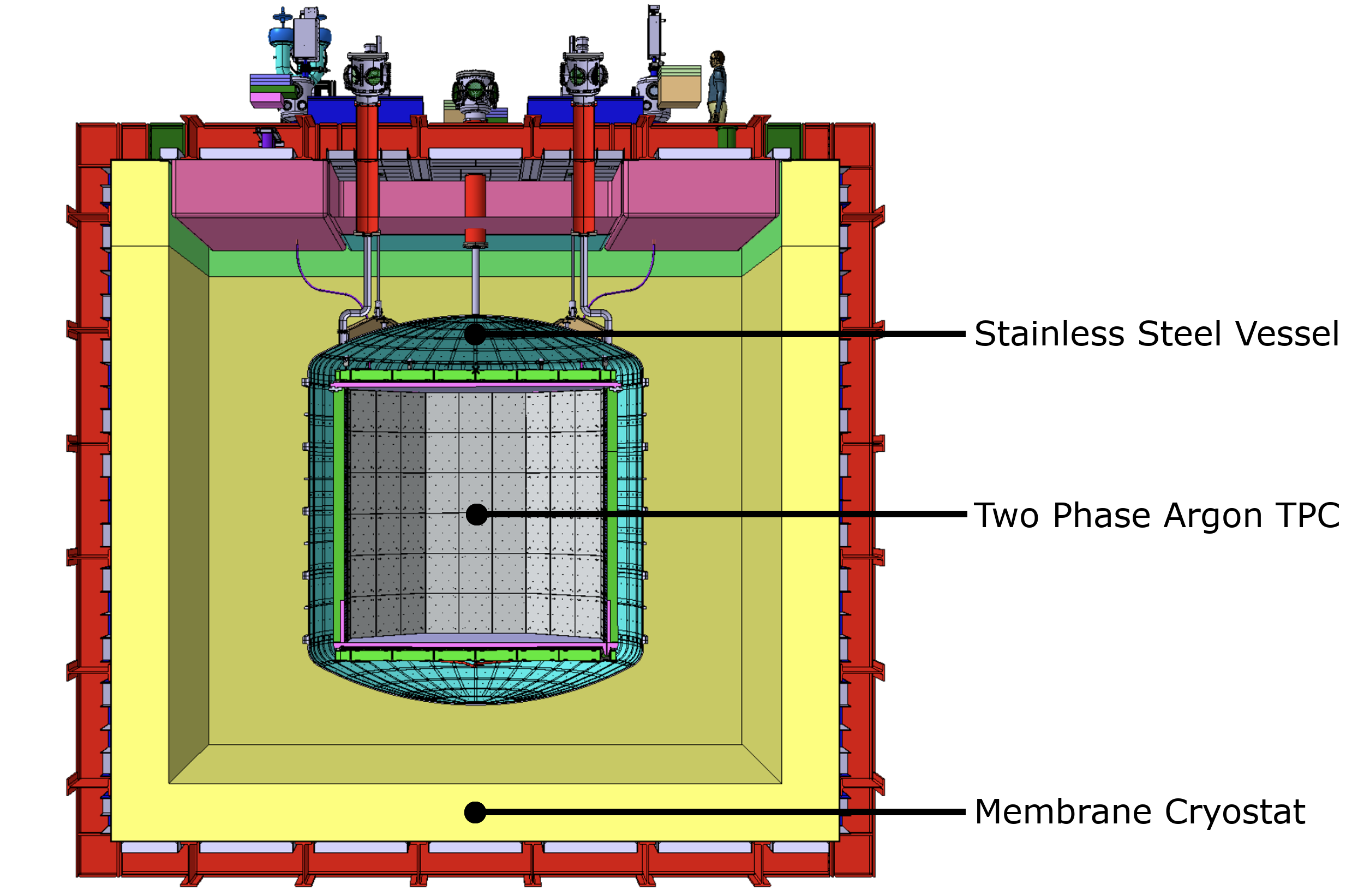}
    \caption{Section of the DarkSide-20k detector assembly. Inside the cryostat lies the inner detector, composed of the TPC and the inner veto system, which are contained within a stainless steel vessel filled with radiopure, underground-extracted argon. Atmospheric liquid argon is used instead in the main cryostat, which is also instrumented with photosensors to serve as muon veto.}
    \label{fig:ds20k_and_veto}
\end{figure}

DarkSide-20k (DS-20k) will feature a dual-phase argon TPC (Figure~\ref{fig:ds20k_and_veto}) with a total argon mass of \SI{50}{\tonne} and maximum sensitivity in the DM mass range of \SI{1}{\GeV/c^2} to \SI{10}{\TeV/c^2}. 
Thanks to the versatility of the dual-phase argon TPC technology, DS-20k will be competitive across a wide range of WIMP masses, spanning both the \(\mathcal{O}(10)\,\mathrm{GeV}/c^2\) to \(\mathcal{O}(100)\,\mathrm{TeV}/c^2\) region, where the most stringent constraints are currently set by xenon-based experiments, and the \SI{1}{\text{-}}\SI{10}{\GeV/c^2} low-mass regime. In the latter case, argon-based detectors are intrinsically favored over xenon due to the lighter nuclear target, which provides a more efficient kinematic matching for low-mass dark matter scattering \cite{DarkSide-20k:2024yfq}.

DarkSide-20k is formed by a set of nested detectors comprising two anti-coincidence (veto) systems that envelope a central dual-phase argon TPC. The outer veto is composed of an instrumented volume of liquid argon housed within the experiment main cryostat, a massive ProtoDUNE-like membrane cryostat~\cite{Montanari:2015pxz} with an internal volume of approximately \SI{600}{\cubic\meter}. 

At the center of the outer veto volume is a stainless steel vessel that contains both the TPC and the inner veto system. The vessel is thermally coupled to the surrounding liquid argon of the outer veto, while being filled with low-radioactivity underground-extracted argon (UAr). The use of UAr, which also constitutes the active target of the TPC, is essential to suppress the radioactive background from the \isotope[39]{Ar} isotope, naturally abundant in atmospheric argon (AAr)~\cite{WARP:2006nsa}. The \isotope[39]{Ar} activity of UAr has been measured by DarkSide-50 to be approximately 1,400 times lower than that of AAr~\cite{DarkSide:2015cqb}.

The DS-20k TPC has the shape of a regular octagonal prism, with an inscribed-circle diameter and a height of \SI{350}{\cm}. It is capped at the top and bottom by two transparent acrylic (PMMA) windows coated with a thin conductive polymer PEDOT:PSS (Clevios\texttrademark), which act as electrodes for the electric field in the drift region (active volume) and the amplification region. The electric field drives ionization electrons from the interaction point toward a thin gaseous argon layer at the top of the detector, where they generate secondary scintillation light (S2 signal). The volume of UAr located outside the TPC but inside the stainless steel vessel is instrumented and functions as the inner veto of the experiment.

The primary and secondary scintillation light are read out by Silicon Photomultipliers (SiPMs), selected as the photon detection technology for DS-20k due to their inherently low radioactivity~\cite{Baudis:2018pdv}, strong performance~\cite{Renker:2006ay}, and suitability for scalable manufacturing processes. The SiPMs utilized in DS-20k were specifically developed for cryogenic operation in collaboration with Fondazione Bruno Kessler~\cite{Gola:2019idb}.

\begin{figure}[t]
    \centering
        \subfloat[\centering]{\includegraphics[height=80pt]{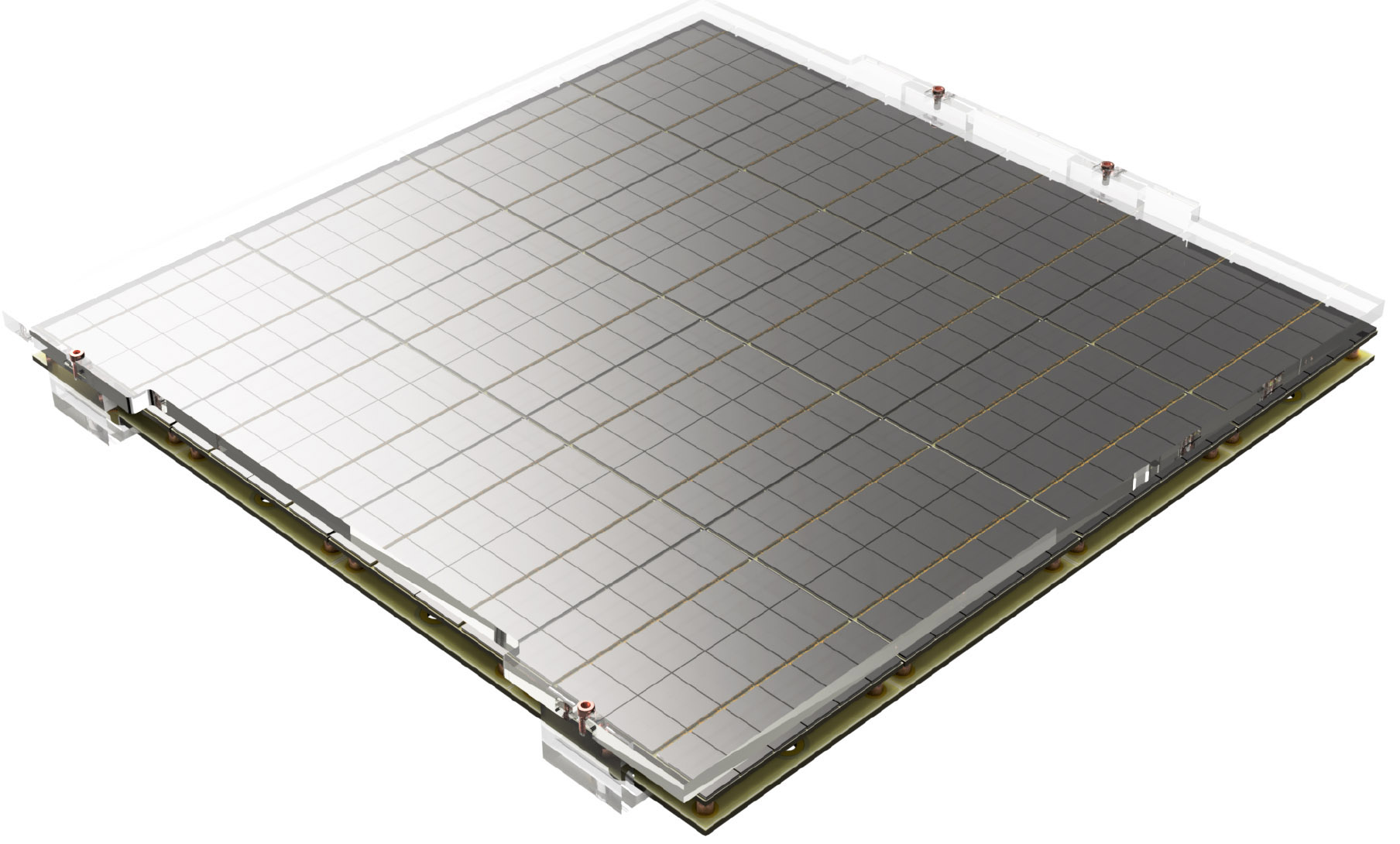}}
        \qquad
    \subfloat[\centering]{\includegraphics[height=80pt]{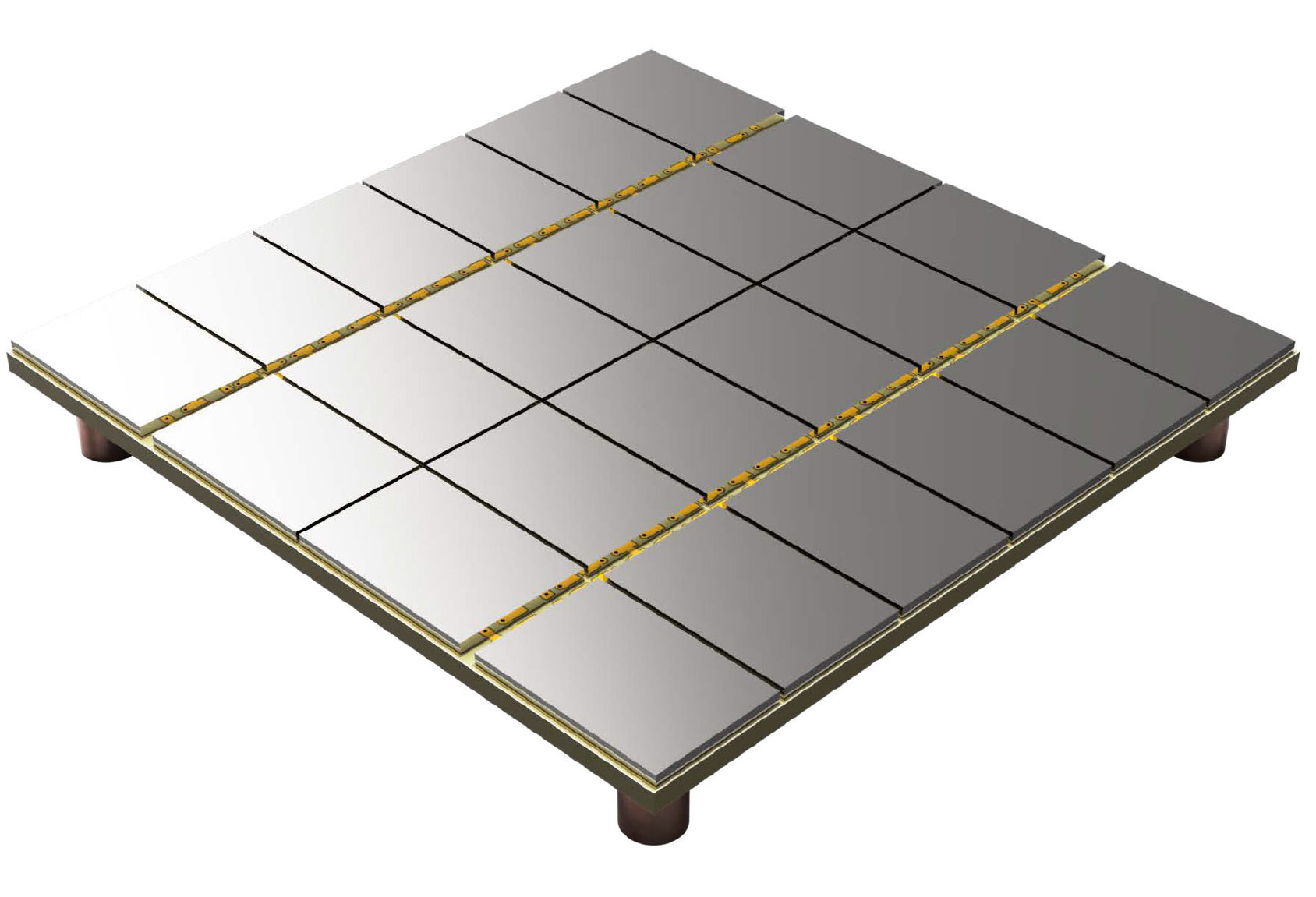}}

    \caption{\textit{(a)} Picture of the fully assembled PDU, with a total surface of $20\times20$~\unit{\cm\squared}. The transparent cover is for protection during transport and handling. \textit{(b)} Top view of a $5\times5$~\unit{\cm\squared} SiPM tile module; 16 tiles are assembled together to form a PDU.}
    \label{fig:pdupdm}
\end{figure}

To instrument the large-scale optical system of DS-20k, the Collaboration developed a modular, cryogenic, SiPM-based photon detector with a total area of $20\times20\,\text{cm}^2$, named the Photon Detector Unit (PDU). Each PDU provides four analog readout channels, each channel covering $100\,\text{cm}^2$ of active SiPM area, and features several stages of amplification and analog summing of SiPM signals to achieve large detection area per channel while maintaining single-photon resolution. Figure~\ref{fig:pdupdm} shows the PDU assembly.

For the DarkSide-20k TPC, a total of 528 PDUs will be installed to form the top and bot optical planes, located beyond the acrylic windows and providing a total of 2,112 readout channels. An additional 160 PDUs are distributed on the surfaces of the inner and outer veto systems to provide light readout for these systems. 

\section{DarkSide Proto-0}

While each of the novel technologies of DS-20k has been individually tested and validated in controlled laboratory environments, their behavior within a dual-phase TPC has not yet been verified. The DarkSide Proto-0 (Proto-0) experiment was therefore conceived as a small-scale analogue of DS-20k, featuring a \SI{7}{\kg} dual-phase TPC and operated at INFN Naples.
Proto-0 is designed to study the integration, signal formation, and optical readout of DS-20k. In addition, it aims to characterize the physics of electroluminescence, namely the extraction of ionization electrons into the gas phase and their conversion into light in dual-phase TPCs, providing feedback on the optimal geometry for DS-20k.

\subsection{Detector, readout and data acquisition}
The TPC is housed in a \SI{300}{L} cryostat that forms part of an integrated cryogenic system inspired by the conceptual design of the DS-20k cryogenics, developed and commissioned at INFN Naples.
The TPC’s main body is built from high-transparency PMMA, with four PMMA panels forming the walls and enclosing a square internal volume of about \SI{10}{\liter}. The active volume is delimited at the top by a wire grid with a \SI{3.0}{\mm} pitch, on the bottom by a PMMA slab, and on the sides by four reflective panels made of high reflectivity ESR foils coated with TPB, the latter used as wavelength shifter to convert the UV scintillation light of argon to the optical spectra, where the SiPMs are most sensitive.

Above the grid, a PMMA ``diving bell'' encloses the region where argon gas accumulates to form the pocket used for generating the secondary ionization signal. The inner top surface of the diving bell, as well as the inner surface of the bottom PMMA slab, are coated with a \SI{10}{\nm} thin-film of Clevios to serve as transparent electrodes for the TPC, offering a transparency of approximately $99\%$ to optical light. The Clevios electrodes are themselves coated with TPB. 

The optical readout is provided by two PDUs identical to those used in DS-20k, whose signals are processed by a differential acquisition board fully integrated into a MIDAS-based control and acquisition system. The setup also includes a laser calibration setup that delivers light from an external pulsed laser to the TPC through cryogenic optical fibers for frequent monitoring and characterization of the PDUs response.

The event waveform is baseline-subtracted channel-by-channel and integrated in a fixed time window around the prompt scintillation pulse; the total charge is converted to photoelectrons from calibration constants obtained from dedicated runs where the single-photoelectron response of the SiPMs is measured.

The initial operation of the detector was performed in single-phase mode, without a gas pocket. 
This configuration provides a clean environment to characterize the fundamental response of the photosensors to scintillation light and to establish the detector baseline \gls{ly}. The results obtained in this mode serve as a reference for dual-phase operation and are reported in this paper.

\section{Photosensor characterization}

Throughout the data-taking campaign, the performance of the two PDUs was monitored with dedicated daily calibration runs. 
These runs provide, on a channel-by-channel basis, the single-photoelectron charge, baseline noise, signal-to-noise ratio, and the photoelectron multiplicity distribution. 
This procedure allows tracking gain and noise variations over month-long timescales, providing a continuous assessment of detector stability under cryogenic operating conditions.

A key parameter derived from calibration data is the duplication coefficient \(k_{\mathrm{dup}}\), which quantifies the contribution of correlated avalanches in the SiPMs. 
Optical cross-talk and afterpulsing can produce additional avalanches following a primary one, resulting in departures from a purely Poisson photoelectron counting statistics. 
We model the peak occupancies obtained from a multi-gaussian fit to the charge (integral) spectrum using a compound-Poisson description~\cite{Vinogradov:2009xxx}, where each primary avalanche produces a geometrically distributed number of secondary avalanches with probability \(p\). 
In this framework,
\[
k_{\mathrm{dup}} = \frac{p}{1-p},
\]
which corresponds to the mean number of secondary avalanches per primary avalanche, and the mean observed multiplicity is \(\langle n\rangle=\mu(1+k_{\mathrm{dup}})\), where \(\mu\) is the mean number of primary photoelectrons in the absence of correlated noise. 
The value of \(k_{\mathrm{dup}}\) is extracted independently for each channel and run and then averaged over channels and over the full campaign.

Biasing the SiPMs with an overvoltage of \SI{7}{V}, we obtain \(k_{\mathrm{dup}} = 0.445 \pm 0.006\), stable within uncertainties over the full dataset.

\section{Single Phase Characterization}
The detector light yield is defined as
\[
  \mathrm{LY} \equiv \frac{N_{\mathrm{PE}}}{E},
\]
where \(N_{\mathrm{PE}}\) is the mean detected photoelectrons and \(E\) the deposited energy. 
We correct for correlated avalanches using the duplication factor \kdup\ measured from daily calibration runs.
In the following, \(\mathrm{LY}\) denotes the raw (observed) light yield, while \(\mathrm{LY}_{\mathrm{corr}}=\mathrm{LY}/(1+\kdup)\).

\begin{figure}[t]
    \centering
    \subfloat[\centering]{
        \includegraphics[width=0.46\textwidth]{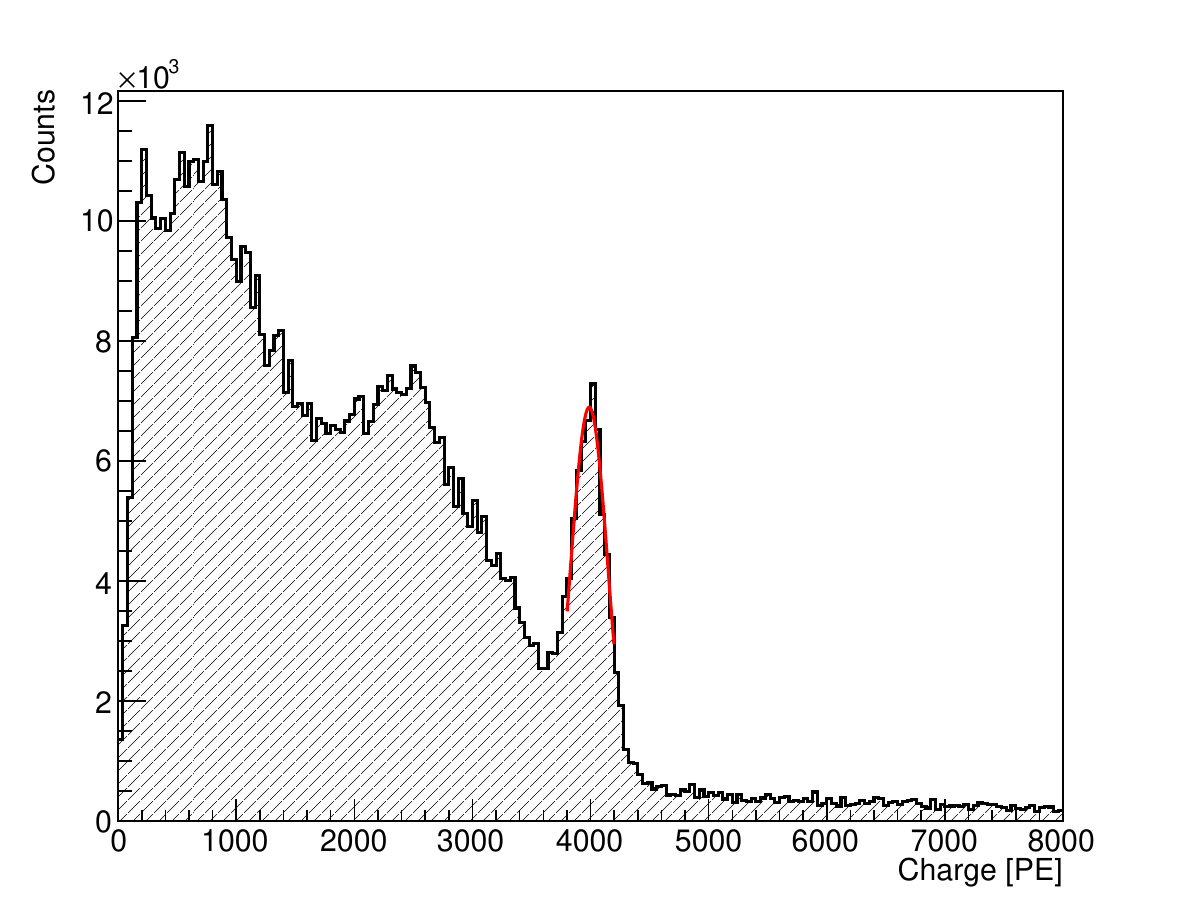}
    }
    \hfill
    \subfloat[\centering]{
        \includegraphics[width=0.46\textwidth]{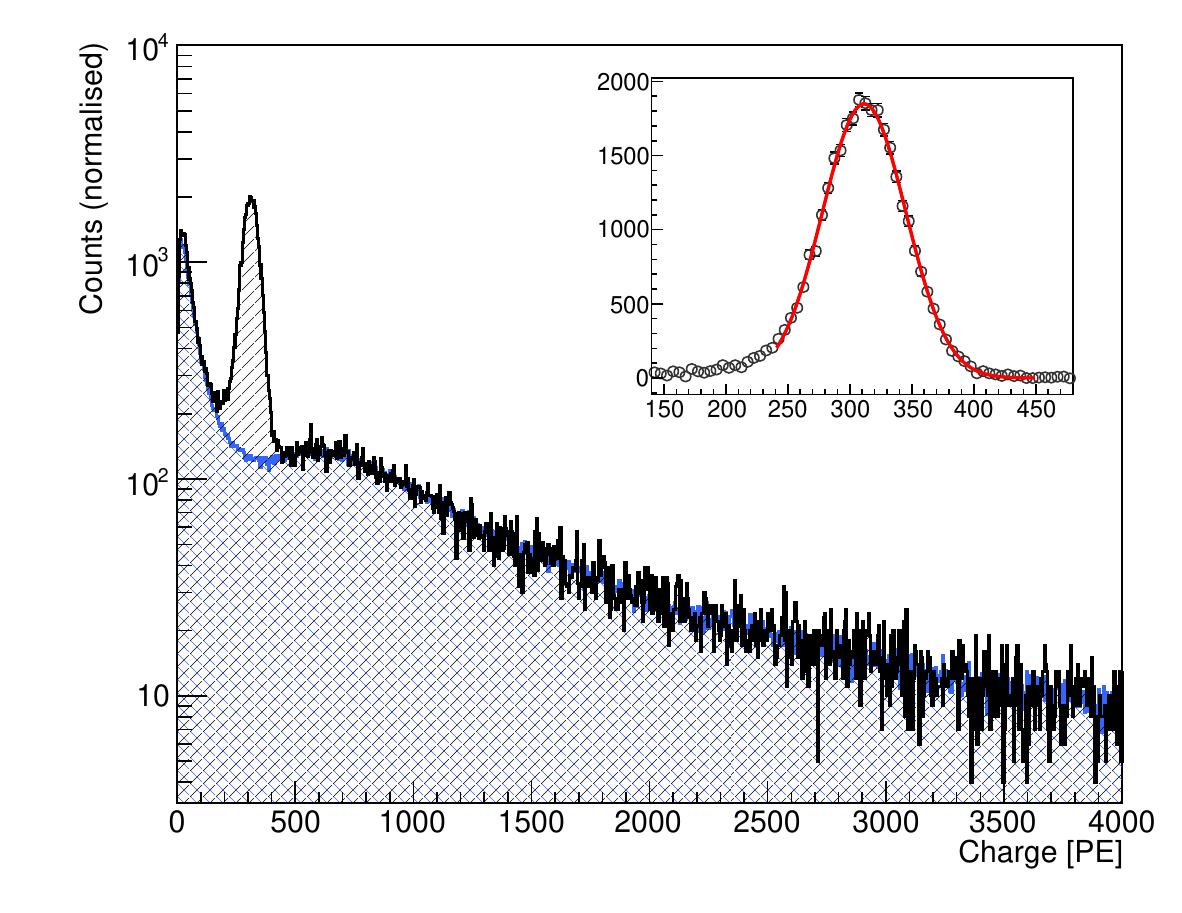}
    }
    \caption[Single-phase spectra in Proto-0]{
    \textit{(a)} Single phase (S1 only) spectrum in PE from a \SI{16}{kBq} \isotope[22]{Na} source at \SI{7}{\V} overvoltage, acquired in coincidence with an external liquid-scintillator detector. The red curve shows a gaussian fit to the \SI{511}{keV} full-absorption feature.
    \textit{(b)} Single phase (S1 only) spectrum in PE after \isotope[83m]{Kr} injection. The black histogram shows data, while the blue shaded histogram indicates background component used for subtraction. The inset shows the background-subtracted \SI{41.5}{keV} peak with a gaussian fit (red).}
    \label{fig:performance:singlephase_spectra}
\end{figure}

The high-energy response was characterized using a \isotope[22]{Na} source at \SI{7}{\V} of overvoltage. 
Events were selected through a coincidence between an external liquid-scintillator detector and a majority trigger on the TPC channels. 
The \SI{511}{keV} feature (Figure ~\ref{fig:performance:singlephase_spectra}a) does not correspond to a purely mono-energetic photoelectric deposition: at this energy, interactions in liquid argon are dominated by Compton scattering, with a Compton-to-photoelectric ratio of nearly 200:1~\cite{NIST_XCOM}. 
The inferred light yield therefore represents an effective response to full-energy absorption following one or more Compton scatters.

Averaging over several runs taken on different days, we obtain
\[
  \gls{ly}=\SI{7.7(1)}{PE\per\keV}, \qquad
  \gls{ly}_{\text{corr}}=\SI{5.3(1)}{PE\per\keV},
\]
with a resolution at \SI{511}{keV} of \(\mathrm{FWHM}/\mu=\SI{9.5}{\percent}\) and stability of the peak of the order of 1\% over repeated measurements.

The low-energy calibration was performed using internally injected \isotope[83m]{Kr}, produced from the decay of \isotope[83]{Rb}. 
The two transitions at \SI{32.1}{keV} and \SI{9.4}{keV} sum to \SI{41.5}{keV} and, given the scintillation and SiPM pulse time scales, are detected as a single pulse. 
Figure~\ref{fig:performance:singlephase_spectra}b shows the krypton spectrum after uniform diffusion in the detector and background subtraction.

From a gaussian fit to the \SI{41.5}{keV} peak we obtain a mean value of \SI{311(1)}{PE}, corresponding to
\[
  \gls{ly}=\SI{7.5\pm0.1}{PE\per\keV}, \qquad
  \gls{ly}_{\text{corr}}=\SI{5.2\pm0.1}{PE\per\keV},
\]
with an energy resolution of \SI{25.2}{\percent}. 
The peak position was stable within the 1\% level over repeated measurements.

\section{Conclusions}
Proto-0 is a small-scale liquid-argon TPC operated at INFN Naples, conceived to validate in a realistic detector environment key DarkSide-20k technologies, with particular emphasis on the integration and long-term performance of SiPM-based photon detectors, and on the study of charge extraction and electroluminescence in dual-phase operation.

We reported the commissioning and single-phase performance of Proto-0. In single-phase operation we measured an uncorrected scintillation light yield of \(\sim\SI{7.5}{PE\per\keV}\) using two radioactive sources, \isotope[22]{Na} and \isotope[83m]{Kr}, with peak positions stable at the percent level.

Dual-phase operation has already been achieved in Proto-0. Dedicated analyses of extraction, electroluminescence and S2 performance, together with extended PDU monitoring in dual-phase conditions, are ongoing and will be presented in a dedicated publication.

\bibliographystyle{JHEP}
\bibliography{pd24.bib}
\end{document}

%% file: acronyms.tex
\newacronym{ds20k}{\mbox{DS-20k}}{\mbox{DarkSide-20k}}
\newacronym{proto}{\mbox{Proto-0}}{DarkSide \mbox{Proto-0}}
\newacronym{ds50}{\mbox{DS-50}}{\mbox{DarkSide-50}}
\newacronym{lar}{LAr}{Liquid Argon}
\newacronym[
  longplural={Time Projection Chambers},
  shortplural={TPCs}
]{tpc}{TPC}{Time Projection Chamber}
\newacronym{pdu}{PDU}{Photon Detection Unit}
\newacronym{pmma}{PMMA}{Polymethyl Methacrylate}
\newacronym{ptf}{PTF}{Photodetector Test Facility}
\newacronym{tpb}{TPB}{Tetraphenylbutadiene}
\newacronym{gpm}{GPM}{Gas Pocket level Meter}
\newacronym{vuv}{VUV}{Vacuum Ultra Violet}
\newacronym{pde}{PDE}{Photon Detection Efficiency}
\newacronym[
  longplural={Silicon PhotoMultipliers},
  shortplural={SiPMs}
]{sipm}{SiPM}{Silicon PhotoMultiplier}
\newacronym{esr}{ESR}{Enhanced Specular Reflector}
\newacronym{fr}{FR}{First Ring}
\newacronym{hv}{HV}{High Voltage}
\newacronym{rtd}{RTD}{Resistance Temperature Detector}
\newacronym{mft}{MFT}{Motion Feedthrough}
\newacronym{noa}{NOA}{Nuova Officina Assergi}
\newacronym{ln}{LN}{Liquid Nitrogen}
\newacronym{gn}{GN}{Gaseous Nitrogen}
\newacronym{mfc}{MFC}{Mass-Flow Controller}
\newacronym{pid}{PID}{Proportional–Integral–Derivative control}
\newacronym{crio}{cRIO}{CompactRIO}
\newacronym{pandid}{P\&ID}{Piping \& Instrumentation Diagram}
\newacronym{al300}{AL300}{AL300 Condenser}
\newacronym{ped}{PED}{Pressure Equipment Directive}
\newacronym{spe}{SPE}{Single Photoelectron}
\newacronym{ptfe}{PTFE}{Polytetrafluoroethylene}
\newacronym{pe}{PE}{Photo-electron}
\newacronym{ov}{OV}{Overvoltage}
\newacronym{pcr}{PCR}{Pulse Count Rate}
\newacronym{roi}{ROI}{Region-of-interest}
\newacronym{dcr}{DCR}{Dark Count Rate}
\newacronym{snr}{SNR}{Signal-to-Noise Ratio}
\newacronym{ly}{LY}{Light Yield}
\newacronym{red}{ReD}{Recoil Directionality}
\newacronym{aar}{AAr}{Atmospheric Argon}
\newacronym{mockup}{Mockup}{DarkSide-Mockup}
\newacronym{uar}{UAr}{Underground Argon}
\newacronym{cmb}{CMB}{Cosmic Microwave Background}
\newacronym{psd}{PSD}{Pulse Shape Discrimination}
\newacronym[
  longplural={Weakly Interacting Massive Particles},
  shortplural={WIMPs}
]{wimp}{WIMP}{Weakly Interacting Massive Particle}
\newacronym[
  longplural={Electron Recoils},
  shortplural={ERs}
]{er}{ER}{Electron Recoil}
\newacronym[
  longplural={Nuclear Recoils},
  shortplural={NRs}
]{nr}{NR}{Nuclear Recoil}
\newacronym{lambdacdm}{$\Lambda$CDM}{$\Lambda$ Cold Dark Matter}
\newacronym{lsv}{LSV}{Liquid Scintillator neutron Veto}
\newacronym{wcd}{WCD}{Water Cherenkov muon veto Detector}
\newacronym{gadmc}{GADMC}{Global Argon Dark Matter Collaboration}
\newacronym{sec}{SEC}{Single Electron Candidates}
\newacronym{tba}{TBA}{Top-Bottom Asymmetry}
\newacronym{el}{EL}{Electroluminescence}
\newacronym{ap}{AP}{Afterpulsing}
\newacronym{ct}{CT}{Crosstalk}
\newacronym{fbk}{FBK}{Fondazione Bruno Kessler}
\newacronym[
  longplural={Single-Photon Avalanche Diodes},
  shortplural={SPADs}
]{spad}{SPAD}{Single-Photon Avalanche Diode}
\newacronym[
  longplural={Photomultiplier Tubes},
  shortplural={PMTs}
]{pmt}{PMT}{Photomultiplier Tube}
\newacronym{tia}{TIA}{Transimpedance Amplifier}
\newacronym{id}{ID}{Inner Detector}
\newacronym{iv}{IV}{Inner Veto}

\newcommand{\mkdup}{k_{\mathrm{dup}}}
\newcommand{\kdup}{\ensuremath{\mkdup}}

